\documentclass[prd,aps, twocolumn, amsfonts,nofootinbib,longbibliography,notitlepage]{revtex4-1}

\usepackage{graphicx}
\usepackage[dvipsnames]{xcolor}
\usepackage{rotating}
\usepackage{amsmath,amssymb,graphics,amsthm,isomath}
\usepackage{amsfonts,dsfont,mathtools}
\usepackage{bbm}
\usepackage{bm}
\usepackage{array}
\newcolumntype{P}[1]{>{\centering\arraybackslash}p{#1}}
\usepackage{multirow}

\usepackage[colorlinks=true, urlcolor=violet, linkcolor=blue, citecolor=red, hyperindex=true, linktocpage=true]{hyperref}
\usepackage[capitalise,compress]{cleveref}

\allowdisplaybreaks

\numberwithin{thm}{section}

\renewcommand{\thesection}{\arabic{section}}
\renewcommand{\thesubsection}{\thesection.\arabic{subsection}}

\makeatletter
\renewcommand{\p@subsection}{}
\renewcommand{\p@subsubsection}{}
\makeatother

\usepackage{xcolor}
\usepackage{mathtools}

\newcommand\bea{\begin{eqnarray}}
\newcommand\eea{\end{eqnarray}}
\newcommand\be{\begin{equation}}
\newcommand\ee{\end{equation}}
\newcommand\bes{\begin{subequations}}
\newcommand\ees{\end{subequations}}
\newcommand\bed{\begin{displaymath}}
\newcommand\eed{\end{displaymath}}
\newcommand\beal{\begin{aligned}}
\newcommand\eeal{\end{aligned}}
\newcommand\bew{\begin{widetext}}
\newcommand\eew{\end{widetext}}
\newcommand\beit{\begin{itemize}}
\newcommand\eeit{\end{itemize}}
\def\bea{\begin{array}}
\def\eea{\end{array}}
\newcommand\been{\begin{enumerate}}
\newcommand\eeen{\end{enumerate}}

\usepackage{verbatim}

\usepackage{dsfont}

\begin{document}

\title{Kinematically constrained vortex dynamics in charge density waves}

\author{Marvin Qi}
\email{marvin.qi@colorado.edu}
\affiliation{Department of Physics and Center for Theory of Quantum Matter, University of Colorado, Boulder CO 80309, USA}

\author{Andrew Lucas}
\email{andrew.j.lucas@colorado.edu}
\affiliation{Department of Physics and Center for Theory of Quantum Matter, University of Colorado, Boulder CO 80309, USA}

\date{\today}

\begin{abstract}
We build a minimal model of dissipative vortex dynamics in two spatial dimensions, subject to a kinematic constraint: dipole conservation.  The additional conservation law implies anomalously slow decay rates for vortices.  We argue that this model of vortex dynamics is relevant for a broad range of time scales during a quench into a uniaxial charge density wave state.  Our predictions are consistent with recent experiments on  uniaxial charge density wave formation in $\mathrm{LaTe}_3$.
\end{abstract}

\maketitle

%\tableofcontents

\section{Introduction}
The dynamics of systems with multipolar symmetries and more general kinematic constraints have been the subject of intense study in recent years. Much of the interest in such systems derives from their natural relation to the study of fractons, which are quasiparticle excitations in many-body systems with restricted mobility \cite{HermeleNandkishoreReview, PretkoChenYouReview, gromov2022fracton, VijayHaahFu_2016, Pretko_subdimensional, Pretko_generalizedEM}. These mobility restrictions often originate from multipolar symmetries or gauged versions thereof \cite{Gromov_2019, Bulmash_2018, dipolarbosehubbard_2022}.  Restricted mobility of microscopic degrees of freedom can lead to many observable consequences in dynamics, including ergodicity breaking \cite{PhysRevB.101.214205, Pozderac_2023}, Hilbert space fragmentation \cite{PhysRevX.9.021003, PhysRevB.101.174204, PhysRevX.10.011047, Moudgalya_2021, PhysRevX.12.011050} and subdiffusive hydrodynamics \cite{GLN_2020, PhysRevLett.125.245303, breakdownhydro_2022, Glorioso_2023, Guo_2022, PhysRevB.100.214301, PhysRevResearch.3.043186, Osborne_2022, PhysRevE.103.022142, PhysRevResearch.2.033129, PhysRevLett.125.245303}.

Given the unusual nature of the symmetries involved in fractonic theories, it is often challenging to realize the dynamical phenomena discovered above directly in experiment.  One exception is the emergence of dipole conservation in tilted optical lattices \cite{elmer, Scherg_2021, kohlert2021experimental}.  Spin liquids and frustrated magnetism \cite{PhysRevLett.124.127203, Hering_2021, Yan_2022, PhysRevLett.128.227201} may also give rise to similar physics, though a conclusive experimental demonstration has not been found yet.  The most ``conventional" realization of such unusual symmetries in nature is in elastic solids:  via fracton-elasticity duality \cite{fractonelasticity_2018, Pretko_2019, Nguyen_2020, fractonsvectorgaugetheory, huang2023chernsimons}, the charges and dipoles of a rank-two tensor gauge theory are mapped to disclination and dislocation defects of a two-dimensional crystal. The disclination is immobile in isolation while the dislocation can only move along its Burgers vector; these mobility constraints are shared respectively by the charge and dipole of the rank-two tensor gauge theory. Similar mobility constraints apply to defects of two-dimensional smectic liquid crystals  \cite{Radzihovsky_2020, Zhai_2021}.

The purpose of this work is to show that in a rather related setting -- the formation of (uniaxial) charge density waves (CDW) -- emergent and approximate mobility constraints can have striking dynamical signatures that are experimentally observable. We will see that topological defects (vortices) have an anomalously long lifetime in uniaxial charge density wave formation. More concretely, we write down a minimal model for dissipative vortex dynamics in this setting, incorporating a dipolar kinematic constraint on vortex motion orthogonal to the density wave direction. 
 Numerical simulations and analytical theory demonstrate that dissipative vortex dynamics of such constrained vortices is qualitatively different from usual dissipative vortex dynamics \cite{ambegaokar,PhysRevB.23.3483, rotatinghelium}, which has been realized in e.g. thin films of superfluid helium \cite{vortex_helium_films}. 
 
 Our predictions are quantitatively consistent with recent ultrafast experiments on $\mathrm{LaTe}_3$ \cite{orenstein2023subdiffusive}, which revealed anomalously slow subdiffusive vortex decay, incompatible with the existing Ginzburg-Landau model of uniaxial density wave formation.  Hence, this work reveals a promising new avenue for studying ``fractonic" dynamical universality classes in quantum materials.

\section{Vortex dynamics in charge density waves}

Topological defects of an order parameter naturally form when a system undergoes a quench from a disordered to an ordered phase \cite{Kibble_1976, Zurek_1985}. Relaxation toward the equilibrium steady state proceeds via annihilation of the topological defects. In a two dimensional uniaxial charge density wave, the topological defects are vortices, which correspond physically to dislocations of the CDW.

\begin{figure}
    \centering
    \includegraphics[width=\columnwidth]{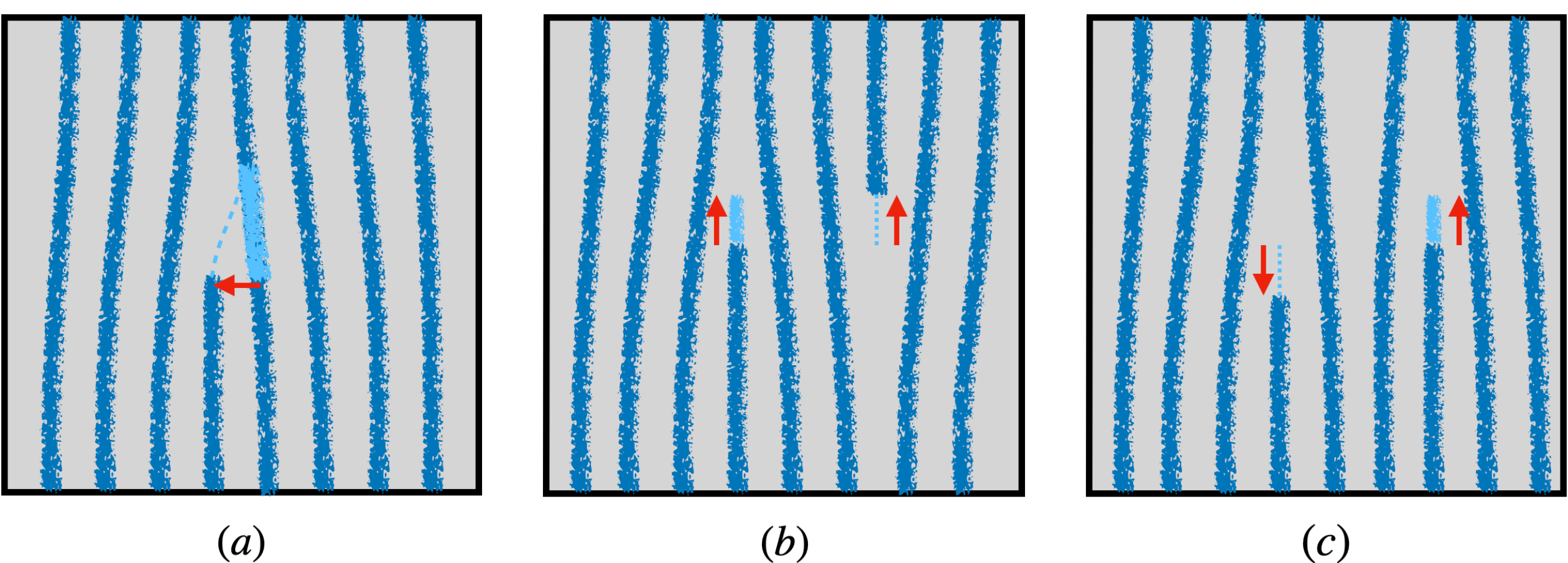}
    \caption{Kinematic constraints of dislocations of the charge density wave. Column $(a)$ shows a local process which translates the vortex freely in the $\hat{x}$ direction. On the other hand, motion of same (resp. opposite) sign vortices along the $\hat{y}$ direction only occurs in pairs, as depicted in $(b)$ (resp. $(c)$). The pair process conserves dipole moment along the $\hat{y}$ direction. } 
    \label{fig:cdwvortex}
\end{figure}

The \emph{equilibrium} properties of the transition into a uniaxial CDW are commonly described using the same Ginzburg-Landau (GL) theory describing the superfluid-insulator transition. However, we argue following \cite{leosmectic} that the standard analysis of a dynamical GL theory will incorrectly describe dynamics of CDW vortices. Unlike superfluid vortices, vortices of the CDW are dislocations, which are subject to approximate kinematic constraints. If the layering occurs perpendicular to the $\hat{x}$ axis, then local operators can translate vortices along the $\hat{x}$ direction, as shown in Fig. \ref{fig:cdwvortex}$(a)$. Motion along $\hat{y}$ requires a nonlocal operator which translates the CDW layer, however, because translating the vortex in this direction requires \emph{adding or removing charge}, which violates local charge conservation if the CDW is in its ground state. Hence, the simplest moves in the $\hat{y}$ direction is of a pair of vortices: see Fig. \ref{fig:cdwvortex}$(b)-(c)$. Such processes leave the $\hat{y}$ dipole moment of the vortices unchanged. 

At finite temperature, we expect a very small density of mobile charged degrees of freedom thermally fluctuating on top of the CDW, which will give a single vortex a small mobility in the $\hat{y}$ direction.  In this Letter,  we will focus on dynamics at short time scales, where this process can be neglected.

\section{The model} \label{sec:vortexmodel}
We now develop a minimal model for vortex dynamics subject to the constraint above.  The degrees of freedom are the positions $\mathbf{r}^\alpha = (x^\alpha, y^\alpha)$ of the $N$ vortices.  Starting with the dissipationless component, we anticipate that this can be described by conventional point-vortex dynamics: after all, such dynamics \emph{already} conserves dipole \cite{gromov_vortices}.  The dissipationless dynamics is moreover Hamiltonian, if we define Poisson brackets
\begin{equation} \label{eq:xypb}
    \{ x^\alpha, y^\beta \} = \frac{1}{\Gamma_\alpha} \delta_{\alpha \beta}.
\end{equation}
Here $\Gamma_\alpha$ is the vorticity of the $\alpha$-th vortex. Note that we do not sum over repeated indices. This can equivalently be written as 
\begin{equation} \label{eq:pb}
    \{r^\alpha_i, r^\beta_j \} = \frac{1}{\Gamma_\alpha} \delta_{\alpha \beta} \epsilon_{ij}.
\end{equation}
The vortices interact via a logarithmic potential, so the Hamiltonian is (in dimensionless units)
\begin{equation} \label{eq:logpotential}
    \mathcal{H} = -\sum_{\alpha < \beta} \Gamma_\alpha \Gamma_\beta \text{ log}(|\mathbf{r}_\alpha - \mathbf{r}_\beta|).
\end{equation}

The corresponding Hamiltonian equations of motion are 
\begin{equation} \label{eq:vortexeom}
\begin{aligned}
    \dot{x}^\alpha &= \{ x, \mathcal{H} \} = \; \; \frac{1}{\Gamma_\alpha} \frac{\partial H}{\partial y^\alpha} \\
    \dot{y}^\alpha &= \{ y, \mathcal{H} \} = -\frac{1}{\Gamma_\alpha} \frac{\partial H}{\partial x^\alpha}
\end{aligned}
\end{equation}

In this setting, dipole conservation is a consequence of translation invariance. Indeed, the Poisson brackets \eqref{eq:xypb} mean that $\Gamma_\alpha y^\alpha$ plays the role of ``momentum" of the $\alpha$-th vortex in the $\hat{x}$ direction, and similarly for $-\Gamma_\alpha x^\alpha$ in the $\hat{y}$ direction. The total dipole moments are therefore identified with the generators of translation, whose conservation follows from translation invariance of $\mathcal{H}$.

The dipole conservation can be seen in the exactly solvable two-body dynamics of vortices. Pairs with equal vorticity travel in a circular orbit around their center of mass, while pairs of opposite vorticity move in a straight line perpendicular to their dipole moment; in each case dipole moment is conserved \cite{gromov_vortices}.

We now turn to the effects of dissipation. The standard model for dissipative dynamics of point vortices is 
\begin{equation} \label{eq:standardvortexdissipation}
\begin{aligned}
    \dot{x}^\alpha &= \frac{1}{\Gamma_\alpha} \frac{\partial H}{\partial y^\alpha} - \gamma \frac{\partial H}{\partial x^\alpha} \\
    \dot{y}^\alpha &= -\frac{1}{\Gamma_\alpha} \frac{\partial H}{\partial x^\alpha} - \gamma \frac{\partial H}{\partial y^\alpha}
\end{aligned}.
\end{equation}
where $\gamma$ term is the mutual friction responsible for dissipation. Note, however, that it breaks the conservation of dipole moment. 

Indeed, one can see the effect of $\gamma$ in the two-body dynamics of vortices. It causes same-sign vortices to drift apart, and opposite-sign vortices to approach each other and collide in finite time; dipole moment conservation is violated in the latter case.

\begin{figure}
    \centering
    \includegraphics[width=\columnwidth]{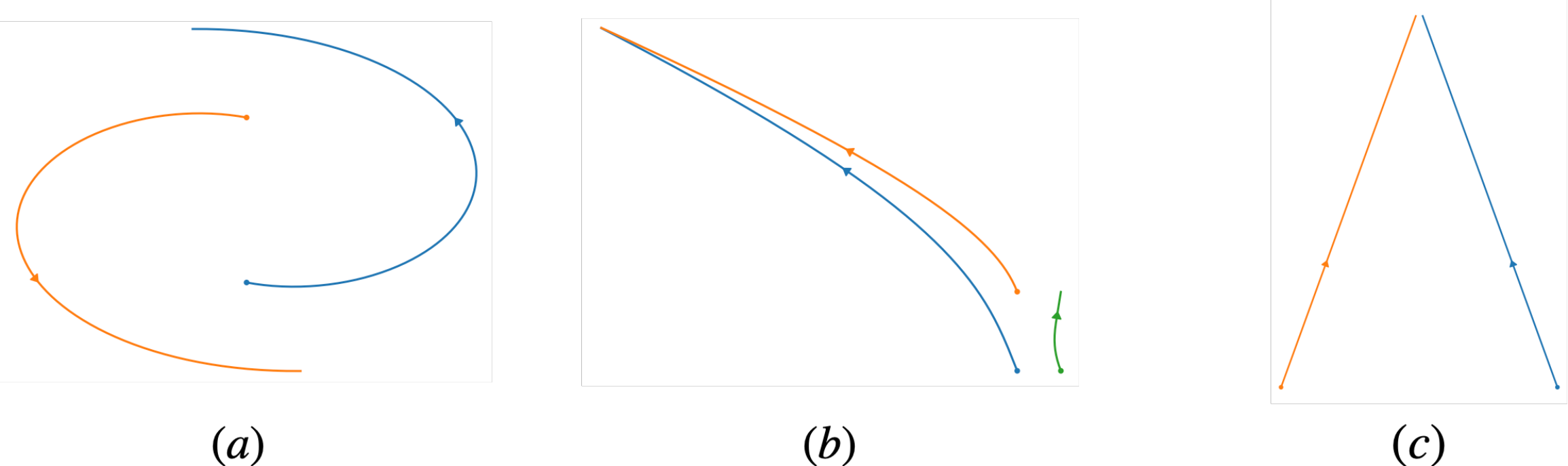}
    \caption{Few body dissipative dynamics of kinematically constrained vortices. $(a)$ Two same-sign vortices will tend to spiral away from each other preserving dipole moment. $(b)$ The minimal three-body process by which two vortices can annihilate. The change in dipole moment is offset by motion of the third vortex. $(c)$ When only the $y$ dipole moment is conserved, two vortices can annihilate if their $y$ dipole moment exactly vanishes.  
    }
    \label{fig:fewvortexdynamics}
\end{figure}

A minimal model for dissipative vortex dynamics which conserve both components of the dipole moment is (see Appendix \ref{sec:eftreview} for a derivation)
\begin{equation} \label{eq:dipoledissipativeeom}
\begin{aligned}
    \dot{x}^\alpha &= \frac{1}{\Gamma_\alpha} \frac{\partial H}{\partial y^\alpha} - \gamma' \frac{\tilde{f}_\alpha}{\Gamma_\alpha^2} \frac{\partial H}{\partial x^\alpha} + \gamma' \sum_\beta \frac{f_{\alpha \beta}}{\Gamma_\alpha \Gamma_\beta} \frac{\partial H}{\partial x^\beta}\\
    \dot{y}^\alpha &= -\frac{1}{\Gamma_\alpha} \frac{\partial H}{\partial x^\alpha} - \gamma' \frac{\tilde{f}_\alpha}{\Gamma_\alpha^2} \frac{\partial H}{\partial y^\alpha} + \gamma' \sum_\beta \frac{f_{\alpha \beta}}{\Gamma_\alpha \Gamma_\beta} \frac{\partial H}{\partial y^\beta}
\end{aligned}.
\end{equation}
where $f_{\alpha \beta} \coloneqq f(|\mathbf{r}^\alpha - \mathbf{r}^\beta|)$ is a function which depends only on the distance between vortices $\alpha$ and $\beta$. The function $f_{\alpha \beta}$ is not constrained by the EFT; we choose $f_{\alpha \beta} = |\mathbf{r}^\alpha - \mathbf{r}^\beta|^{-1}$. When this dipole-conserving dissipative term is included, two vortices of opposite sign can approach each other and annihilate in the presence of a nearby third vortex. The motion of the third vortex compensates for the change of dipole moment caused by the annihilation of the two vortices, leaving total dipole moment unchanged. This process is depicted in Fig. \ref{fig:fewvortexdynamics}(b). 
We find, however, that for our choice of $f_{\alpha \beta}$, if the initial positions are sufficiently far apart then a vortex pair can simply escape off to infinity without annihilating.  

\begin{figure}
    \centering
    \includegraphics[width=0.75\columnwidth]{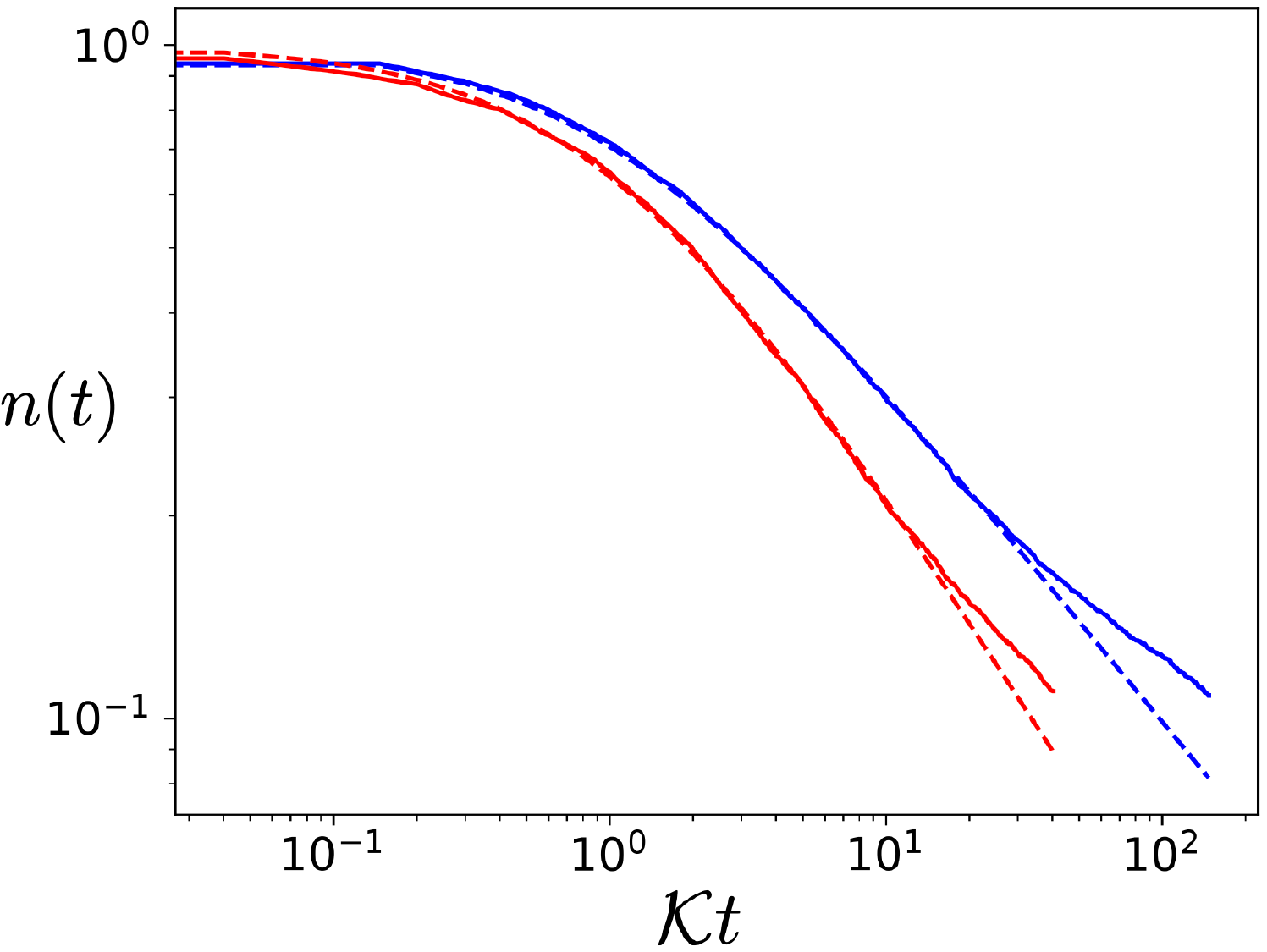}
    \caption{Average number of vortices $\langle n(t) \rangle$ normalized as a fraction of the initial number of vortices. The blue and red curves show $\langle n(t) \rangle$ for vortices which conserve dipole moment in both directions and only the $\hat{x}$ direction, respectively. The dashed lines show their corresponding fits to the function $1/(1+\mathcal{K} t)^{\alpha}$, with $\alpha_{\text{dipole}} = 0.50 \pm 0.01$ and $\alpha_{x-\text{dipole}} = 0.65 \pm 0.02$.  Note that the $x$ axis is scaled to $\mathcal{K}t$; $\mathcal{K}$ is a fit parameter which is different for the two systems.}
    \label{fig:n(t)}
\end{figure}

We numerically simulate the $N$-body dynamics of dipole-conserving vortices given by the equations of motion \eqref{eq:dipoledissipativeeom}. For the initial conditions we randomly sample $N$ points uniformly from a box of size $L \times L$, and randomly assign vorticities $\pm 1$ to each.  Dissipation causes vortices to come together and annihilate in finite time. Vortex annihilation is implemented in the simulation by manually removing pairs of opposite-sign vortices when their distance decreases below a cutoff $\epsilon$. We plot the average surviving fraction of vortices $\langle n(t) \rangle$ as a function of (rescaled) time in Fig. \ref{fig:n(t)} in blue for $N = 400$ and $L = 80$. This vortex relaxation process is well-described at early to intermediate times by a function of the form $(1 + \mathcal{K}t)^{- \alpha_\text{dipole}}$ with $\alpha_{\text{dipole}} = 0.50 \pm 0.01$, shown in orange in Fig. \ref{fig:n(t)}. At late times, vortex annihilation occurs much more slowly; this can be attributed to the annihilation process ``freezing out" at sufficiently low density as alluded to above. This is qualitatively similar to the breakdown of thermalization in dipole-conserving systems found in \cite{PhysRevB.101.214205, Pozderac_2023}. 

Vortices of the CDW order parameter (approximately) conserve dipole moment only along the layering axis, and are unconstrained transversely. Assuming layering along the $y$ axis, this anisotropic mobility constraint is implemented by including the $i=x$ terms of \eqref{eq:dissipative} and the $i=y$ terms of \eqref{eq:dipoledissipative} in the EFT Lagrangian. The resulting equations of motion are 
\begin{equation} \label{eq:ydipoledissipativeeom}
\begin{aligned}
    \dot{x}^\alpha &= \frac{1}{\Gamma_\alpha} \frac{\partial H}{\partial y^\alpha} - \gamma \frac{\partial H}{\partial x^\alpha} \\
    \dot{y}^\alpha &= -\frac{1}{\Gamma_\alpha} \frac{\partial H}{\partial x^\alpha} - \gamma' \frac{\tilde{f}_\alpha}{\Gamma_\alpha^2} \frac{\partial H}{\partial y^\alpha} + \gamma' \sum_\beta \frac{f_{\alpha \beta}}{\Gamma_\alpha \Gamma_\beta} \frac{\partial H}{\partial y^\beta}
\end{aligned}.
\end{equation}

Since motion along the $\hat{x}$ axis is unconstrained, the three-body annihilation process of Fig. \ref{fig:fewvortexdynamics}(b) is no longer the only process by which vortices can annihilate. A pair of vortices can annihilate via the process of Fig. \ref{fig:fewvortexdynamics}(c) if the pair has net zero $y$-dipole moment. However, such a process is fine-tuned, as it requires the $y$-dipole moment to vanish exactly. Nevertheless, we expect that the dynamics should proceed more quickly than in the isotropic dipole-conserving case, as it is less constrained.  

We follow the same procedure to simulate the $N$-body dynamics of \eqref{eq:ydipoledissipativeeom} with vortex annihilation. The average surviving fraction of vortices is plotted in green in Fig. \ref{fig:n(t)}, for $N=400$ and $L = 80$. The red dashed line shows the fit at early to intermediate times to $(1+\mathcal{K}t)^{-\alpha_{y-\text{dipole}}}$, with $\alpha_{y\text{-dipole}} = 0.65 \pm 0.02$. The relation $\alpha_{y-\text{dipole}} \gtrsim \alpha_{\text{dipole}}$ is consistent with faster dynamics as a consequence of fewer kinematic constraints.

In an isotropic theory, we can successfully estimate $\alpha$ for both the dipole-conserving and non-conserving vortex dynamics based on evaluating the two-point correlator for a conserved (vortex) density $n$: \begin{equation}
  \langle n(t) n(0)\rangle \sim  \int \mathrm{d}k_x \mathrm{d}k_y \mathrm{e}^{-D_x k_x^{a_x}t - D_y k_y^{a_y}t} \sim t^{-\alpha^*}, 
\end{equation}with $\alpha^* = 1/a_x + 1/a_y$, using $a=4$ for dipole-conserving \cite{GLN_2020,leosmectic} and $a=2$ for dipole-non-conserving dissipative dynamics.  If only $y$-dipole is conserved, this argument then suggests that $\alpha = \frac{1}{2}+\frac{1}{4}=0.75$. The numerically observed $\alpha_{y\text{-dipole}} = 0.65$ is not consistent with this estimate.  This suggests that the dynamical universality class observed is not fully captured by ``mean field" scaling.

\section{Reaction-diffusion dynamics}

To further justify our observation of $\alpha=0.5$ scaling observed when both components of dipole moment are conserved, we employ the theory of stochastic reaction-diffusion equations \cite{1983JChPh..78.2642T, OVCHINNIKOV1978215, PhysRevLett.52.955, BurlatskyOshanin, PhysRevLett.62.694.2, cardy_cardy_falkovich_gawedzki_2008}. Our setup, however, contains some complicating features: the vortices experience long-range interactions given by the logarithmic potential \eqref{eq:logpotential}, and their dynamics conserve the total dipole moment (along one or both axes). These features modify the kinetic rate equations and their scaling. 

Moreover, in contrast to an ordinary $A + B \to 0$ reaction-diffusion system, two isolated vortices are unable to annihilate each other, as doing so would change the dipole moment of the system. Rather, two vortices can only annihilate in the presence of another nearby vortex. In this case, the change in dipole moment arising from annilation of the vortex pair is compensated by motion of the nearby vortex. See Fig. \ref{fig:fewvortexdynamics}(b) for an illustration of the minimal process by which a pair of vortices can annihilate while preserving dipole moment in both directions.  Hence, the allowed reactions are of the form $A+B+x\rightarrow x$, for $x=A,B$.

Letting $\rho_{A,B}$ denote the densities of the two species (positive/negative vortices), we then postulate a kinetic rate equation
\begin{equation}
    \frac{\mathrm{d} \rho_A}{\mathrm{d} t} = \frac{\mathrm{d} \rho_B}{\mathrm{d} t} -\mathcal{K} \rho_A \rho_B (\rho_A + \rho_B), 
\end{equation}
Defining the total vortex density $n = \rho_A + \rho_B$ and the (signed) vorticity density $\rho = \rho_A - \rho_B$, we have 
\begin{equation} \label{eq:dipoleconservedK}
\begin{aligned}
    \frac{\mathrm{d} n}{\mathrm{d} t} &= -\frac{\mathcal{K}}{2} (n^2 - \rho^2) n \\
    \frac{\mathrm{d} \rho}{\mathrm{d} t} &= 0
\end{aligned}
\end{equation}
The second equation is the statement that the total charge is a conserved quantity since only opposite-sign vortices can annihilate. When the initial charge density vanishes, \eqref{eq:dipoleconservedK} can be solved to give 
\begin{equation} \label{eq:dipolemeanfield}
    n(t) = (1/n_0^2 + \mathcal{K} t)^{-1/2}
\end{equation}
which is in sharp contrast to the case where no dipole symmetry is imposed. The exponent is in excellent agreement with the numerical results of the isotropic dipole-conserving vortex model of the previous section. 

We now include the effects of spatial fluctuations and long-range interactions in the reaction-diffusion dynamics. The equations of motion are modified to be
\begin{equation} \label{eq:dipoleLRIeom}
\begin{aligned}
    \partial_t n - D \nabla^2 n &= -\frac{\mathcal{K}}{2} [n^2 - \rho^2]n - Q \nabla (\rho \nabla V) \\
    \partial_t \rho - \tilde{D} \nabla^4 \rho &= -Q \nabla ( n \nabla V)
\end{aligned}
\end{equation}
where the potential $V(\mathbf{r}, t)$ is given by 
\begin{equation}
    V(\mathbf{r}, t) = - \int \mathrm{d}^2 \mathbf{r}' \rho(\mathbf{r}', t) \log |\mathbf{r}' - \mathbf{r} | 
\end{equation}
and captures the drift of vortices due to the velocity fields created by the others.  
We have introduced (sub)diffusion coefficients $D$ and $\tilde{D}$ and a coefficent $Q \sim \Gamma^2$ proportional to the square of the vorticity.

Following \cite{LR_reactiondiffusion}, we analyze these equations using a self-consistent approximation where the fluctuations of the total number density $n$ are neglected while the fluctuations of the charge density $\rho$ are kept. In other words, the number density $n(\mathbf{r}, t) = n(t)$ is approximated as a spatially independent function of time. Our goal is to determine the average number density $n(t)$ averaged over an ensemble of initial conditions for $\rho(\mathbf{r}, 0)$ taken to be 
\begin{equation} \label{eq:initialconditions}
\begin{aligned}
    \langle \rho(\mathbf{r}, 0) \rangle &= 0 \\
    \langle \rho(\mathbf{r}_1, 0) \rho(\mathbf{r}_2, 0) \rangle &= n_0^2 \delta^{(2)}(\mathbf{r}_1 - \mathbf{r}_2)
\end{aligned} \; .
\end{equation}
Substituting $n(\mathbf{r}, t) \simeq n(t)$ into the first equation of \eqref{eq:dipoleLRIeom} and averaging over all space and initial conditions, we obtain
\begin{equation} \label{eq:avgdipoleeom}
    \frac{\mathrm{d} n(t)}{\mathrm{d} t} + \frac{\mathcal{K}}{2} n(t)^3 = \frac{\mathcal{K}}{2} n(t) \int \mathrm{d}^2 \mathbf{r} \;  \langle \rho(\mathbf{r},t)^2 \rangle  .
\end{equation}
with $\langle \ldots \rangle$ denoting the average over initial conditions. Fourier transforming $\rho(\mathbf{r}, t)$ in space, the second equation of \eqref{eq:dipoleLRIeom} becomes
\begin{equation}
    \partial_t \rho(\mathbf{k}, t) + D k^4 = -Q n(t) \rho(\mathbf{k}, t)
\end{equation}
away from $k=0$ and $\partial_t \rho(\mathbf{0}, t) = 0$ at $k=0$. This equation can be solved exactly to give
\begin{equation}
    \rho(\mathbf{k}, t) = \rho(\mathbf{k},0) \exp \left( -\tilde{D} k^4 t - Q \int_0^t \mathrm{d}t' n(t') \right).
\end{equation}
Substituting the solution $\rho(\mathbf{k},t)$ into \eqref{eq:avgdipoleeom} and performing the average over initial conditions  \eqref{eq:initialconditions}, the equation of motion for $n(t)$ becomes 
\begin{equation} \label{eq:dipole}
\begin{aligned}
    \frac{\mathrm{d} n}{\mathrm{d} t} + \frac{\mathcal{K}}{2} n^3 &= \frac{\mathcal{K}}{2} n \int \mathrm{d}^2 \mathbf{k} \exp \left( -2 \tilde{D} k^4 t - 2 Q \int_0^t \mathrm{d} t' n(t') \right) \\
    &\simeq \mathcal{K} n \exp \left(-2Q \int_0^t \mathrm{d} t' n(t') \right) \frac{1}{\sqrt{2 D t}}
\end{aligned}.
\end{equation}
This equation reduces to the mean field solution \eqref{eq:dipolemeanfield} when the right hand side can be ignored. The self-consistency of the mean field approximation can be determined as follows. Substituting the mean field solution \eqref{eq:dipolemeanfield} into \label{eq:dipole}, the terms on the left hand side decay as $t^{-3/2}$, while the right hand side decays superpolynomially as $t^{-1}\exp(-\sqrt{t})$.
% \begin{equation} \label{eq:RHSdecay}
%   t^{-1} \exp \left(-2\frac{Q}{\mathcal{K}} \sqrt{2 \mathcal{K} t} \right) .
% \end{equation}
We see that the mean field solution is valid asymptotically for any nonzero $Q$. This justifies the agreement between the numerical simulations of the isotropic dipole-conserving model and the fit obtained from mean-field reaction-diffusion theory. 

\section{Comparison to experiment on $\mathrm{LaTe}_3$}

A recent experiment, which served as motivation for this work, observed anomalously slow dynamics of vortex strings in $\mathrm{LaTe_3}$ \cite{orenstein2023subdiffusive}, whose behavior was not explained by a Ginzburg-Landau type phenomenological theory. 
%The authors instead proposed that subdiffusive dynamics were a consequence of the restricted mobility of the dislocations of the CDW.
The experiment pulses the CDW state of $\mathrm{LaTe_3}$ to photoinduce vortex strings and uses ultrafast x-ray scattering to resolve the time dynamics of the resulting non-equilibrium state. X-ray scattering measures the structure factor $S(\mathbf{k}, t)$, which takes the form 
\begin{equation}
    S(\mathbf{k},t) = g(t) F[k L(t)]
\end{equation}
where $F$ is a universal function and $L(t) \sim t^\beta$ corresponds to the average distance between topological defects. They find a scaling exponent $\beta = 0.29$, which is inconsistent with the standard result $\beta=0.5$ for diffusive vortex decay in superfluid thin films \cite{PhysRevB.23.3483} or superfluid ultracold atomic gases \cite{PhysRevA.90.063627}.

Because vortex strings are a codimension-two defect, the density of defects scales as $n(t) \sim L(t)^{-2} \sim t^{-2\beta}$. While our model of uniaxial CDW vortex relaxation is strictly two-dimensional, it may capture similar phenomena to the experiment, so long as the bending of vortex strings does not play a qualitatively important role in the dynamics (note that 2d vs. 3d does not change $\beta$ in ordinary superfluids). Above, we computed $n(t)$ within our model of CDW vortex relaxation and found  $n(t) \sim t^{-\alpha}$ with $\alpha = 0.65 \pm 0.02$. This yields $\beta = 0.325 \pm 0.01$, which is quite close to the experimentally observed value.  Importantly, our computation of $\beta$ goes beyond the Ginzburg-Landau type phenomenological theory, which produces $\beta = 1/4$ and $\beta=1/2$ for conserved and non-conserved dipole moments, respectively.

\section{Conclusion}
We have constructed a minimal dissipative model of kinematically-constrained vortices, relevant to dynamics over a large range of time scales in uniaxial CDWs. While our isotropic dipole-conserving model agrees well with simple mean-field-theoretic arguments, the experimentally relevant model where only one component of dipole is conserved exhibits anomalous exponents that are close to those observed in a recent experiment on $\mathrm{LaTe}_3$ \cite{orenstein2023subdiffusive}.  Our results therefore provide a quantitative theory for how experimentally-observed subdiffusive dynamics of solid-state defects follows from emergent mobility restrictions, with direct implications for experiment. Generalizing our theory to broader settings, including the constrained dynamics of topological defects in three dimensional charge density waves, remains an interesting open problem that is -- at least in principle -- straightforwardly tackled using the methods described here.

%Our results demonstrate the utility of effective field theory as a quantitative tool to study universal aspects of dynamical systems.

\section*{Acknowledgements}
We thank Leo Radzihovsky, Mariano Trigo and Gal Orenstein for useful discussions.  This work was supported by the U.S. Department of Energy, Office of Science, Basic Energy Sciences under Award number DESC0014415 (MQ), the Alfred P. Sloan Foundation through Grant FG-2020-13795 (AL), and the National Science Foundation through CAREER Grant DMR-2145544 (AL).

\appendix
\renewcommand{\thesubsection}{\thesection.\arabic{subsection}}
\section{Review of effective field theory}
\label{sec:eftreview}
\subsection{Formalism}
\label{sec:formalism}
We review the general effective field theory (EFT) for stochastic dynamics following \cite{Guo_2022}. Let $\mathbf{q} = (q_1, \ldots, q_M)$ be the ``slow" degrees of freedom that we keep track of in the EFT. To begin, we assume the existence of a stationary distribution
\begin{equation} \label{eq:Peq}
    P_{eq} \propto e^{-\Phi[\mathbf{q}]},
\end{equation}
where $\Phi[\mathbf{q}]$ is a function(al) of the degrees of freedom $\mathbf{q}$. The EFT describes the evolution of the probability distribution $P[\mathbf{q},t]$ as it relaxes toward the stationary distribution \eqref{eq:Peq}. The EFT is encoded by an action of the form 
\begin{equation}
    S = \int \mathrm{d}t \sum_a  \left[ \pi_a \partial_t q_a - H(\pi_a, q_a) \right] 
\end{equation}
where $\pi_a$ are ``noise" variables conjugate to $q_a$. We will refer to $H(\pi_a, q_a)$ as the EFT Hamiltonian. 

There are several general conditions that $H(\pi_a, q_a)$ must obey for the action to represent a sensible EFT. We present the general principles that the dynamics must obey and the corresponding consequences for $H(\pi_a, q_a)$; for a derivation of the latter from the former, see \cite{Guo_2022}. First, conservation of probability enforces 
\begin{equation}
    H(\pi_a = 0, q_a) = 0. 
\end{equation}
This ensures that all terms of $H$ have at least one factor of $\pi_a$. Stationarity of $P_{eq}$ places another constraint on the allowed form of $H(\pi_a, q_a)$. Define the generalized chemical potentials 
\begin{equation}
    \mu_a = \frac{\partial \Phi}{\partial q_a}.
\end{equation}
Stationarity of $P_{eq}$ implies that 
\begin{equation}
    H(\pi_a = i \mu_a, q_a) = 0.
\end{equation}
Finally, we assume that the degrees of freedom $\mathbf{q}$ undergo stochastic dynamics whose fluctuations are bounded. This enforces the condition 
\begin{equation} \label{eq:noisepositivity}
    \text{Im}(H) \leq 0.
\end{equation}
Finally, we may require that the dynamics respect a version of time-reversal symmetry. 
Let $\mathbb{T}$ denote the time-reversal operation, which sends $t \to -t$ and $\mathbf{q} \to \mathbb{T}(\mathbf{q})$. 
On the conjugate noise variables $\pi_a$, time reversal acts as
\begin{equation}
    \pi_a \to -\mathbb{T}(\pi_a) + i \mu_a.
\end{equation}
Note that this is a $\mathbb{Z}_2$ transformation. 

The conditions on the EFT Hamiltonian $H(\pi_a, q_a)$ significantly constrain the terms which can appear. At the quadratic level, the Hamiltonian must take the form 
\begin{equation} \label{eq:eftgeneralhamiltonian}
    H(\pi_a, q_a) = \sum_{ab} i \pi_a \mathcal{Q}_{ab} (\pi_b - i \mu_b)
\end{equation}
for matrix $\mathcal{Q}_{ab}$. Decomposing $\mathcal{Q}_{ab} = A_{ab} - \frac{1}{2} S_{ab}$ into its symmetric part $-\frac{1}{2}S$ and antisymmetric part $A$, we  see that $S$ must be positive definite due to \eqref{eq:noisepositivity}. Symmetric contributions to $\mathcal{Q}_{ab}$ are dissipative, while antisymmetric contributions are non-dissipative.

We conclude the review of the formalism with a brief discussion of how additional conservation laws can be accounted for in the dynamics. Suppose we would like to enforce that a quantity $F(\mathbf{q})$ is conserved. By Noether's theorem, this is equivalent to enforcing the shift symmetry 
\begin{equation} \label{eq:pishift}
    \pi_a \to \pi_a + \frac{\partial F}{\partial q_a}
\end{equation}
on the EFT Hamiltonian $H(\pi_a, q_a)$. Naturally, enforcing this symmetry leads to constraints on the allowed terms which can appear in the EFT Hamiltonian. We will encounter many examples of conservation laws below. 

\subsection{Example: diffusion}
Let us illustrate how diffusion of a single conserved density can be recovered within this framework. The only degree of freedom we keep track of is $\mathbf{q} = \rho(x)$, which is a conserved density. Its corresponding conjugate field is $\pi(x)$. We take the stationary distribution to be 
\begin{equation}
    \Phi[\rho(x)] =  \int \mathrm{d}^d x \; \frac{1}{2} \chi \rho^2 + \ldots 
\end{equation}
where the terms in $\ldots$ are higher order in $\rho$. The chemical potential to leading order is $\mu = \chi \rho$. In addition to the aforementioned conditions on $H$, we also demand that $Q = \int \rho$ is conserved. Applying the continuum analogue of \eqref{eq:pishift}, we require that the EFT Hamiltonian is invariant under
\begin{equation}
    \pi \to \pi + c(t).
\end{equation}
To quadratic order $H$ takes the form \eqref{eq:eftgeneralhamiltonian}; the above symmetry fixes it to be
\begin{equation}
    H(\pi, \rho) = -i \sigma \nabla \pi \nabla (\pi - i \mu) + \cdots .
\end{equation}
This term is dissipative since it is a symmetric contribution to \eqref{eq:eftgeneralhamiltonian}. The equation of motion from varying $\pi$ (and then setting $\pi$ to zero) is 
\begin{equation}
    \partial_t \rho - \nabla (D \nabla \rho) = 0
\end{equation}
which is the diffusion equation with $D = \chi \sigma$. 

\subsection{Example: Hamiltonian mechanics}
The formalism can also capture Hamiltonian mechanics in the presence of dissipation. We take our degrees of freedom to be $\mathbf{q} = (x_1, \ldots, x_M, p_1, \ldots, p_M) = (\boldsymbol{x}, \boldsymbol{p})$, with the usual Poisson brackets
\begin{equation}
\begin{aligned}
    \{ x_i, p_j \} &= \delta_{ij} \\
    \{x_i, x_j \} = \{ p_i, p_j \} &= 0
\end{aligned}.
\end{equation}
We assume that there is a Hamiltonian $\mathcal{H}(x,p)$ (to be distinguished from the $H(\pi_a, q_a)$ of the effective field theory) which generates time evolution in the absence of noise and defines the equilibrium distribution 
\begin{equation}
    P_{eq} \propto e^{-\mathcal{H}(x,p)}.
\end{equation}
Correspondingly the chemical potentials are $\boldsymbol{\mu} = (\partial \mathcal{H} / \partial \boldsymbol{x}, \partial \mathcal{H} / \partial \boldsymbol{p})$. Hamilton's equations in the absence of dissipation can be reproduced from the EFT action 
\begin{equation}
    S = \int \mathrm{d}t \sum_a \pi_a \partial_t q_a - \pi_a \{q_a, q_b\} \mu_b.
\end{equation}
The second term is an antisymmetric contribution to the EFT Hamiltonian \eqref{eq:eftgeneralhamiltonian}, so it is dissipationless as expected. We can add dissipation by including in $H(\pi_a, q_a)$ a term 
\begin{equation} \label{eq:sab_dissipation}
    -i \pi_a S_{ab} (\pi_b - i \mu_b)
\end{equation}
where $S_{ab}$ is a positive definite symmetric matrix. 

Additional conservation laws are implemented by enforcing invariance under \eqref{eq:pishift}. In the absence of dissipation, this is equivalent to the condition 
\begin{equation}
    \{F, \mathcal{H} \} = 0
\end{equation}
for a conserved quantity in Hamiltonian mechanics.

\subsection{Dipole-conserving vortices} \label{sec:dipolevortexeft}
We now derive the model of dipole-conserving vortices with dissipation presented in the main text. Recall that the dynamics without dissipation are characterized by the Poisson brackets \eqref{eq:xypb}, \eqref{eq:pb} and Hamiltonian \eqref{eq:logpotential}. Following the previous subsection, we define the generalized chemical potentials $\mu^\alpha_i = \partial \mathcal{H}/ \partial r^\alpha_i$ and write the dissipationless contribution to the EFT Lagrangian as 
\begin{equation}
\begin{aligned}
    L &= \sum_{\alpha, i} \Bigg( \pi^\alpha_i \partial_t r^\alpha_i - \pi^\alpha_i \sum_{\beta, j} \{r^\alpha_i, r^\beta_j \} \mu^\beta_j \Bigg) \\
      &= \sum_{\alpha, i} \Bigg( \pi^\alpha_i \partial_t r^\alpha_i - \pi^\alpha_i \frac{1}{\Gamma_\alpha} \sum_j \epsilon_{ij} \mu^\alpha_j \Bigg).
\end{aligned}
\end{equation}
The ordinary mutual friction term which appears in dissipative models of vortices is recovered by including 
\begin{equation} \label{eq:dissipative}
    \sum_{i\alpha} -i \frac{\gamma}{2} \pi_i^\alpha \left(  \pi^\alpha_i -i \mu^\alpha_i \right).
\end{equation}
as a term in $H(\pi_a, q_a)$. This is simply \eqref{eq:sab_dissipation} where $S_{ab}$ is diagonal. The resulting equations of motion are given by \eqref{eq:standardvortexdissipation}. 

As noted in the main text, under these dynamics dipole moment is not conserved. The total dipole moment is given by $D_i(\mathbf{q}) = \sum_\alpha \Gamma_\alpha r_i^\alpha$, so conservation of dipole moment corresponds via \eqref{eq:pishift} to invariance under $\pi^\alpha_i \to \pi^\alpha_i + \Gamma_\alpha \delta_{ik}$. It is straightforward to see that the term \eqref{eq:dissipative} is not invariant under this transformation. 

To get a dissipative term which respects dipole conservation, the simplest term at quadratic order is given by 
\begin{equation} \label{eq:dipoledissipative}
    \sum_{i\alpha \beta} -i \frac{\gamma'}{2} f_{\alpha \beta} 
    \left(\frac{\pi_i^\alpha}{\Gamma_\alpha} - \frac{\pi_i^\beta}{\Gamma_\beta} \right) \left( \left(\frac{\pi^\alpha_i - i \mu^\alpha_i}{\Gamma_\alpha} \right) - \left(\frac{\pi^\beta_i - i \mu^\beta_i}{\Gamma_\beta} \right) \right)
\end{equation}
where $f_{\alpha \beta} \coloneqq f(|\mathbf{r}^\alpha - \mathbf{r}^\beta|)$ is a function which depends only on the distance between vortices $\alpha$ and $\beta$. This term is clearly invariant under the transformation $\pi^\alpha_i \to \pi^\alpha + \Gamma_\alpha \delta_{ik}$. The function $f_{\alpha \beta}$ is not constrained by the EFT; we choose $f_{\alpha \beta} = |\mathbf{r}^\alpha - \mathbf{r}^\beta|^{-1}$. While the nonlocality of $f_{\alpha \beta}$ may seem unnatural, we emphasize that any microscopic dynamics preserving dipole moment leads to an effective dissipative term of this form, up to a choice of $f_{\alpha \beta}$. The resulting equations of motion are given in \eqref{eq:dipoledissipativeeom}.

\subsection{On the choice of $f_{\alpha\beta}$}
Given the freedom within the EFT to make different choices of $f_{\alpha \beta}$ it is natural to ask what, if anything, singles out the choice $f_{\alpha \beta} = |\mathbf{r}^\alpha - \mathbf{r}^\beta|^{-1}$. We argue via a simple scaling argument that this is the least nonlocal $f_{\alpha \beta}$ which does not cause generic vortex dipoles to escape to infinity without annihilating. 

Consider a vortex dipole together with a third vortex as in Fig.~\ref{fig:fewvortexdynamics}(b). Let us call the distance between the two vortices comprising the dipole $d$ and the distance between the dipole and the third vortex $R$. When dissipation is absent, an isolated dipole will travel at constant speed perpendicular to its dipole moment, so $R \sim t$. In the presence of dissipation, these vortices will approach each other with speed $\dot{d} \sim -\gamma$ where $\gamma$ is the effective dissipation strength. 
Choosing $f_{\alpha \beta} = |\mathbf{r}^\alpha - \mathbf{r}^\beta|^{-\eta}$, the effective dissipation scales as $\gamma \sim 1/R^\eta$. Altogether, we obtain 
\begin{equation}
    \frac{\mathrm{d}}{\mathrm{d}t}d \sim -\frac{1}{t^\eta}.
\end{equation}
When $\eta > 1$, $d$ asymptotes to a constant as $t \to \infty$, and the vortex dipole escapes to infinity without annihilating; when $\eta < 1$, the dipole annihilates in finite time. When $\eta =1$, the dipole always annihilates eventually, but the time to annihilation is exponential in the initial separation $d_0$. For $\eta=1$ we therefore expect to see the dynamics slow down dramatically when the density drops to a point where inter-vortex separation is $O(1)$ in our dimensionless distance units, which is indeed seen in our numerics.

That the vortex dipole escapes to infinity is not an issue if we perform our calculations using periodic boundary conditions; however, this complicates the problem significantly and requires substantially more computational resources. To avoid this issue we simply choose $f_{\alpha \beta}$ so that dipoles don't escape to infinity, which results in the choice in the main text. 

\section{Review of two-species annihilation} \label{sec:ordinaryRD}
In this appendix we review the reaction-diffusion model governing two-species annihilation, considering the cases with and without long-range interactions. We closely follow the discussion in \cite{LR_reactiondiffusion}. 

Ordinary two-species annihilation processes are governed at the mean-field level by a kinetic rate equation 
\begin{equation}
    \frac{\mathrm{d} \rho_A}{\mathrm{d} t} = 
    \frac{\mathrm{d} \rho_B}{\mathrm{d} t} = - \mathcal{K} \rho_A \rho_B \; ,
\end{equation}
which captures the fact that an annihilation process requires two species to be present at the same location. Let us introduce the number density $n$ and the charge density $\rho$ as
\begin{equation}
\begin{aligned}
    n &= \rho_A + \rho_B \\
    \rho &= \rho_A - \rho_B
\end{aligned} \;\; . 
\end{equation}
In terms of $n$ and $\rho$, the rate equation becomes 
\begin{equation}
\begin{aligned}
    \frac{\mathrm{d} n}{\mathrm{d} t} &= -\frac{\mathcal{K}}{2} (n^2 - \rho^2) \\[0.3em]
    \frac{\mathrm{d} \rho}{\mathrm{d} t} &= 0
\end{aligned}
\end{equation}
where the latter equation shows that charge is conserved. When there is an equal initial density of $\rho_A$ and $\rho_B$, \emph{i.e.} at charge neutrality, the asymptotic behavior of this equation is given by 
\begin{equation} \label{eq:ABmeanfield}
    n(t) \sim (\mathcal{K} t)^{-1}.
\end{equation}
The mean-field description is valid in dimensions above the critical dimension $d_c=4$. Below the critical dimension, diffusive and stochastic effects become important, modifying the long-time behavior to 
\begin{equation}
    \rho(t) \sim (D t)^{-d/4}.
\end{equation}
Here $D$ is the diffusion constant. This behavior and the critical dimension can be derived by considering the long-range interacting case discussed below and taking the limit where the interaction strength vanishes, as was shown in \cite{LR_reactiondiffusion}.  

%\subsection{Long-range interactions}
We now treat the case of an ordinary $A+B \to 0$ reaction-diffusion system with long-range interactions, reviewing the calculation of \cite{LR_reactiondiffusion}. In terms of the number density $n$ and charge density $\rho$, the equations of motion are 
\begin{equation} \label{eq:longrangeeom}
\begin{aligned}
    \partial_t n - D \nabla^2 n &= - \mathcal{K} \left[ n^2 - f^2 \right] - Q \nabla \left( \rho \nabla 
    V \right) \\
    \partial_t \rho - D \nabla^2 \rho &= -Q \nabla \left( n \nabla V \right)
\end{aligned}
\end{equation}
where the potential $V(\mathbf{r}, t)$ is given by 
\begin{equation}
    V(\mathbf{r}, t) = - \int \mathrm{d}^2 \mathbf{r}' \rho(\mathbf{r}', t) \log |\mathbf{r}' - \mathbf{r} | .
\end{equation}
We have introduced a diffusion coefficient $D$ and a coefficent $Q \sim \Gamma^2$ proportional to the square of the vorticity.  The authors in \cite{LR_reactiondiffusion} analyzed these equations using a self-consistent approximation where fluctuations of the total number density $n$ are neglected while fluctuations of the charge density $\rho$ are kept. In other words, the number density $n(\mathbf{r},t) = n(t)$ is taken to be a spatially independent function of time. Our goal will be to determine the number density $n(t)$ averaged over an ensemble of initial conditions for $f(\mathbf{r}, 0)$, which we take to be 
\begin{equation} \label{eq:initialconditionsappendix}
\begin{aligned}
    \langle \rho(\mathbf{r}, 0) \rangle &= 0 \\
    \langle \rho(\mathbf{r}_1, 0) \rho(\mathbf{r}_2, 0) \rangle &= n_0^2 \delta^{(2)}(\mathbf{r}_1 - \mathbf{r}_2)
\end{aligned} \; .
\end{equation}
We will normalize $n_0 = 1$. Approximating $n(\mathbf{r},t) \simeq n(t)$, we average the first equation of \eqref{eq:longrangeeom} over all space and over initial conditions to obtain 
\begin{equation} \label{eq:avgneom}
    \frac{\mathrm{d} n(t)}{\mathrm{d} t} + \mathcal{K} n(t)^2 = \frac{\mathcal{K}}{V} \int \mathrm{d}^2 \mathbf{r} \;  \langle \rho(\mathbf{r},t)^2 \rangle  \;.
\end{equation}
with $\langle \ldots \rangle$ denoting the average over initial conditions. Fourier transforming $\rho(\mathbf{r}, t)$ in space, the second equation of \eqref{eq:longrangeeom} gives 
\begin{equation}
    \partial_t \rho(\mathbf{k},t) + D k^2 \rho(\mathbf{k},t) = -Q n(t) \rho(\mathbf{k},t)
\end{equation}
away from $k=0$ and $\partial_t \rho(\mathbf{0},t) = 0$ at $k=0$. The equation of motion for $\rho(\mathbf{k},t)$ can be solved exactly to yield 
\begin{equation}
    \rho(\mathbf{k}, t) = \rho(\mathbf{k}, 0) \exp \left( -D k^2 t - Q \int_0^t \mathrm{d}t' n(t') \right).
\end{equation}
Substituting the solution $\rho(\mathbf{k},t)$ into \eqref{eq:avgneom} and performing the average \eqref{eq:initialconditions} gives 
\begin{equation}
\begin{aligned}
    \frac{\mathrm{d} n}{\mathrm{d} t} + \mathcal{K} n^2 &= \mathcal{K} \int \mathrm{d}^2 \mathbf{k} \exp \left( -2 D k^2 t - 2 Q \int_0^t \mathrm{d} t' n(t') \right) \\
    &= \mathcal{K} \exp \left(-2Q \int_0^t \mathrm{d} t' n(t') \right) \frac{\pi}{2 D t}
\end{aligned}
\end{equation}
where in the second line we performed the integral over $\mathbf{k}$. The mean field solution \eqref{eq:ABmeanfield} was obtained from ignoring the RHS of the above equation. This is valid if, at long times, terms on the RHS decay more quickly than terms on the LHS. Substituting the mean field solution, we see that the terms on the LHS decay as $t^{-2}$, while the RHS decays as $t^{-\alpha}$ with $\alpha = 2Q/\mathcal{K} + 1$. In other words, the mean field solution is valid for $2Q/\mathcal{K} > 1$. 

Repeating the calculation for general spatial dimension $d$ (while modifying $V$ to be the Coulomb potential in $d$ dimensions), one finds that the RHS decays as $\alpha = d/2 + 2Q/\mathcal{K}$. This gives the critical dimension above which mean field theory is valid as $d_c = 4 \left( 1 - Q/\mathcal{K} \right)$. For a system without long-range interaction $(Q = 0)$ as in the previous subsection, this reproduces the result for the upper dimension $d_c = 4$.

\bibliography{thebib}
\end{document}